\documentclass[draft]{agujournal2019}
\usepackage{url} 
\usepackage{lineno}
\usepackage[inline]{trackchanges} 
\usepackage{soul}
\usepackage{framed}
\usepackage{amsmath}
\usepackage{comment}
\usepackage[none]{hyphenat}
\usepackage[skip=0.333\baselineskip]{caption}
\usepackage{subcaption}
\usepackage{todonotes}
\usepackage{MnSymbol}
\usepackage{booktabs}
\usepackage{float}

\nolinenumbers
\usepackage{amsmath}
\usepackage{mdframed}
\usepackage{todonotes}

\draftfalse

\journalname{JGR: Machine Learning and Computation}

\begin{document}

%
%


\title{Anomalous Diffusion of Tropical Cyclones Observed in Huge Ensembles of Hindcasts}

%
%




\authors{Abdoul R. ZEBA\affil{1,2}${}^{\dagger\ast}$,
William D. Collins\affil{3,2}${}^{\dagger^\ast}$, Ankur Mahesh\affil{2,4}${}^{\ast}$, Boris Bonev\affil{5},  Karthik Kashinath\affil{5}, Thorsten Kurth\affil{5}, Michael S. Pritchard\affil{5,6}, and  Shashank Subramanian\affil{7}}


\affiliation{1}{Criteo AI Lab, Paris, \^{I}le-de-France, France}
\affiliation{2}{Earth and Environmental Sciences, Lawrence Berkeley National Laboratory (LBNL), California, USA}
\affiliation{3}{The School of Environmental Sciences, University of East Anglia, Norwich, UK}
\affiliation{4}{University of California, Berkeley, California, USA}
\affiliation{5}{NVIDIA Corporation, Santa Clara, California, USA}
\affiliation{6}{Department of Earth System Science, University of California, Irvine, USA}
\affiliation{7}{National Energy Research Scientific Computing Center (NERSC), LBNL, Berkeley, California, USA}
\affiliation{\text{\dag}}{These authors contributed equally to this work.}




\correspondingauthor{${}^{\ast}$Abdoul R. ZEBA, Ankur Mahesh, William Collins}{azeba@lbl.gov, ankur.mahesh@berkeley.edu, wdcollins@lbl.gov}



\begin{keypoints}
\item Statistics of tropical cyclones in summer 2023 have been quantified using huge ensembles of hindcasts with a machine learning weather prediction model.
\item The spread of the cyclone tracks with time from cyclone initiation can be treated as a diffusive process.
\item Huge ensembles confirm that cyclones obey anomalous super-diffusion and can approach ballistic limits of passage through the background flow.
\end{keypoints}

%
%

%
%


\begin{abstract}

We examine whether tropical cyclones (TCs) obey ordinary Brownian or anomalous diffusion using a huge ensemble (HENS) of hindcasts for summer 2023.  Anomalous diffusion has been inferred for actual TCs from the fluctuations in their tracks from the shortest paths between the initiation and termination of each cyclone.  We reproduce the same anomalous diffusion power laws connecting spatial position and time using HENS.  In addition, we show that the variance in the position of a single TC across HENS  since initiation follows a scaling law with time that, in some cases, corresponds to ballistic motion of the TC through the background atmospheric flow.  This determination was enabled by the exceptional statistics determined from thousands of plausible yet counterfactual recreations of 34 individual TCs.  HENS consists of 7424 15-day hindcasts initiated from observed atmospheric conditions  each day from June~1, 2023 to August~31, 2023 using the ECMWF ERA5 meteorological reanalysis.  The hindcasts were generated using NVIDIA's Spherical Fourier Neural Operator (SFNO) machine-learning-based weather and climate emulator.  We identify tropical cyclones in HENS using a variant of the Tempest Extremes detection and tracking frameworks for TCs with adjustments to the disposable parameters to minimize the numbers of false positives and negatives relative to the International Best Track Archive for Climate Stewardship (IBTrACS)  records for TCs observed in summer 2023.   We conclude with the implications of our findings for the predictability of TC tracks and landfall locations on lead times of days to weeks.

\end{abstract}

\section*{Plain Language Summary}
\label{sec:PlainLanguage}
We have used a machine learning emulator of the Earth's weather to simulate the hurricanes and tropical cyclones during summer 2023, the second-hottest summer in the last 2000 years.  We have created a huge ensemble of plausible reconstructions of each recorded tropical cyclone from that summer to understand how the uncertainties in projected tracks of these storms grow with time. These uncertainties have significant impacts on the lead times available to prepare coastal communities in the likely paths of these storms.  In general, using these huge ensembles, we find the uncertainty in track projections is appreciably larger than previously estimated. We show that the tropical cyclones can move through the  atmospheric wind fields in their environment in a ballistic manner, with their directions of travel essentially unimpeded by countervailing winds.

%
%

\section{Introduction}
\label{sec:Introduction}
Studies of the diffusive properties of the Earth's atmosphere were initiated by pioneering experiments by \citeauthor{richardson1926} \citeyear{richardson1926} involving weather balloons serving as passive tracers.  Diffusive properties can be characterized by the relationship between the mean-squared displacement of such tracers and the time since the release of the tracers into the fluid flow field of interest.  For ordinary Brownian motion, that relationship is linear, whereas for anomalous diffusion, the relationship is nonlinear.  \citeauthor{richardson1926} established that the weather balloons he released obeyed a class of anomalous diffusion known as super-diffusion, in which the mean-squared displacement among the tracers grows faster with time than when subject to ordinary Brownian motion.  With the advent of comprehensive global meteorological reanalyses, studies of the classes of diffusive motion have been extended to the global atmosphere \cite{Huber2001}.

Anomalous diffusion of tropical cyclones (henceforth TCs) has attracted wide interest in meteorological and dynamical systems studies \cite{Blender1997}. Laboratory analogues of tropical cyclones emulated using columnar vortices in rotating Rayleigh-B\'enard convection also exhibit anomalous diffusion \cite{Noto_Tasaka_Yanagisawa_Murai_2019}.  Realistic observations show that the trajectories of TCs often deviate from classical Brownian motion and, instead, display complex patterns that can be described by anomalous diffusion \cite{Meuel:hal-00708481}. Such deviations come from the interaction of the atmospheric factors: large-scale circulation, local environmental conditions, and intrinsic dynamics of TCs \cite{McCloskey2013, Zehnder2024}. Indeed, studies up to today have described such behaviors using observational data interpreted using the theory of fractional Brownian motion  \cite{BERNIDO20142016,Bernido_2015}. However, deepening our understanding of these phenomena has been limited by the inherent weaknesses of observational data, including sparse sampling, data quality issues, and finite record lengths.  

Two questions serve as the primary motivations for this study: First, do huge ensemble weather-emulation systems reproduce anomalous diffusion as observed for collections of real TCs? Secondly, and more importantly, does HENS provide a way to actually calculate a classical diffusion exponent for TCs using dispersion measures, something not possible with previous observational data? Ensemble systems generate large sample sizes, thereby offering a unique platform for these questions. The current work aims to bridge the gap between theoretical models of diffusion and practical constraints in observational TC analysis by leveraging the spatial and temporal richness of HENS.  
 


\section{Methods}
\label{sec:Methods}
\subsection{Motivation for Studying Summer 2023}
\label{ssec:Motivation}
\begin{figure}
\centering
\includegraphics[width=1\linewidth]{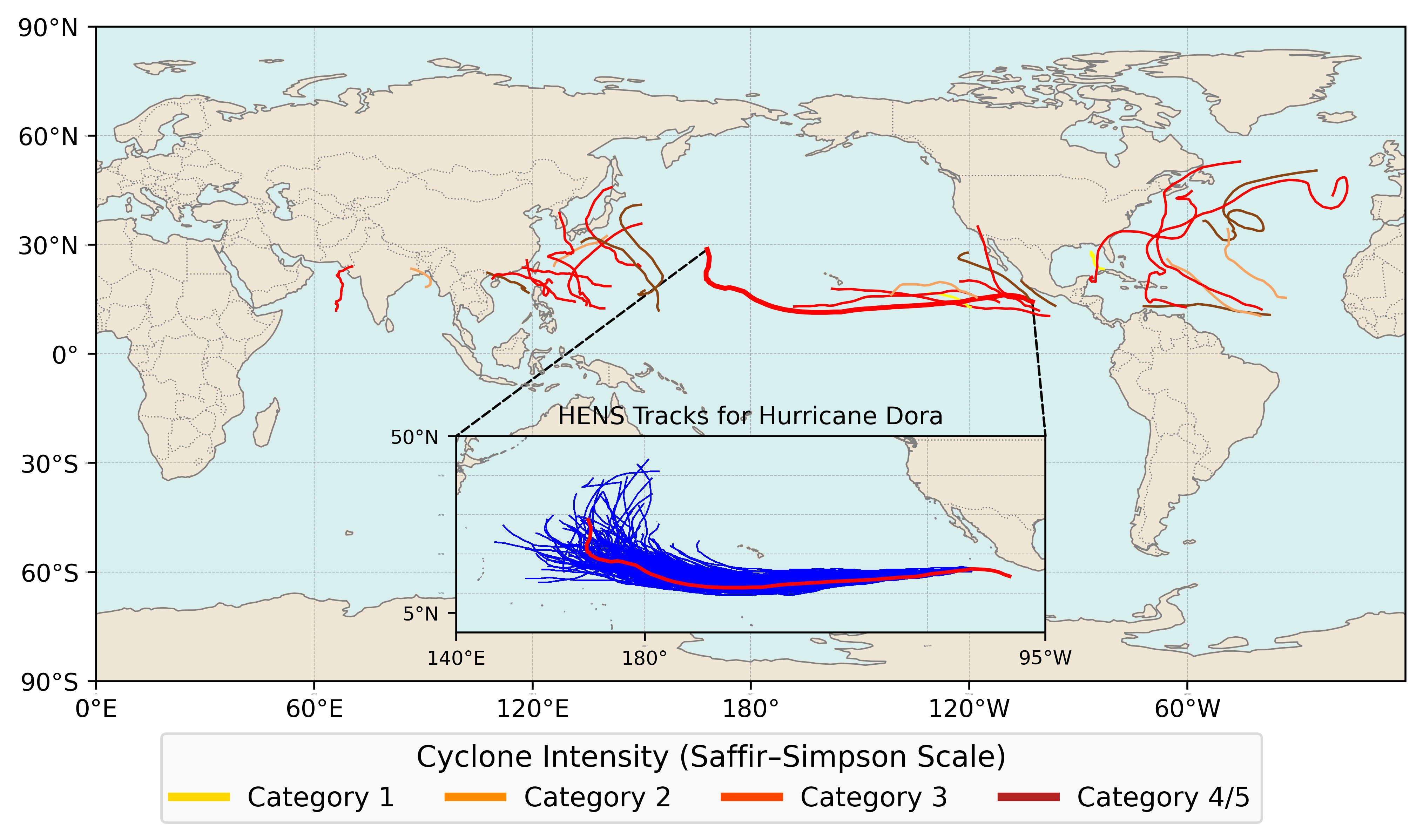}
\caption{\textbf{Tracks of  observed and named tropical cyclones initiated between 1 June and 31 August, 2023.} Tracks are color-coded by the Saffir-Simpson scale \protect\cite{Saffir1973,Simpson1974}\label{fig:TC_Map}. The inset depicts 100 of the 7424 counterfactual tracks for Hurricane Dora generated with HENS. \label{fig:Tracks}}
\end{figure}

We focus this study on the behavior of cyclones during the boreal summer of 2023 (Figure~\ref{fig:TC_Map}), which until 2024 was the hottest summer in the observational record \cite{WMO2023} and probably during the last 2000 years \cite{Esper2024,Hegerl2024}. The temperature anomalies during that summer are well in excess of those predicted by observationally calibrated statistical climate models and, to date, have not been definitely attributed to causal agents \cite{Schmidt2024}.   The annually averaged temperature of 2023 was higher than a climatological baseline during 1850-1900 by 1.43${}^\circ$C  \cite[central estimate]{Forster2024} and warmer by 1.49${}^\circ$C relative to a pre-1700 baseline \cite[central estimate]{Jarvis2024}.  The departures from historical conditions are unprecedented in the quantitative observational record and signal the emergence of a climate with no recent historical analogue \cite{Ripple2023}.  We elected to construct our huge ensemble of hindcasts for this summer in order to understand how several types of extremes, including tropical cyclones but also heatwaves and atmospheric rivers, would respond to these warmer conditions.

\subsection{Overview of HENS}
\label{ssec:HENS}
Several threads motivate research  on Low-Likelihood, High-Impact {LLHI} extreme events such as TCs. First, the most recent Intergovernmental Panel on Climate Change Report has an increased focus on LLHIs while also recognizing that future predictions of their occurrence currently have \emph{low confidence}. Second, the occurrence of recent LLHIs, like the Summer 2021 PNW heatwave, reveals that our abilities to characterize, let alone anticipate, such events are currently incomplete.  

LLHIs challenge the standard climate models that might be used to investigate such issues.  The climate community has recognized that such massive ensembles are needed to characterize high-impact but low-likelihood events \cite{Thompson2017}. Only high-resolution, convection-resolving climate model simulations have demonstrated fidelity in simulating tropical cyclones, but computational costs make it impractical to run the large ensembles of simulations that are necessary to make inferences about the statistics of extremely rare hurricanes like Harvey under this approach.  That said, recent development of ensemble-boosting techniques has shown considerable promise for sampling rare, large, spatially coherent events such as heatwaves \cite{Ragone2017,Ragone2019,Fischer2023}. 

These challenges motivate the application of entirely new methodological approaches -- such as those based on artificial intelligence and machine learning (AI/ML).
We have created a huge ensemble of weather extremes using an ML-based emulator of global numerical weather reanalyses and climate simulations of the recent climate record \cite{Mahesh2024hugeensemblesidesign,Mahesh2024hugeensemblespartii}.  Our goal is to conduct exploratory studies of whether massive ensembles from ML/AI emulators improve the statistical fidelity of LLHI drivers.  

This paper is based upon an ML weather model known as FourCastNet \cite{Kurth2022,Pathak2022,Li2020}, which was trained deterministically based on the {Spherical Fourier Neural Operator (SFNO)} \cite{Bonev2023} architecture.
HENS is an ensemble of hindcasts generated using an initial-condition perturbed variant of SFNO that uses bred vectors and multiple equivalently plausible trained ML model checkpoints to achieve realistic probabilistic calibration
\cite{Mahesh2024hugeensemblespartii}. More precisely, it constitutes a 15-day-long ensemble hindcast of 7,424 ensemble members, initialized at each day from June 1st through August 31st, 2023.

To generate the 7,424 ensemble members for a given day, we initialized FourCastNet using initial conditions from the ERA5 reanalysis dataset on each given day of interest. We then introduced perturbations to these initial conditions based on the bred vector method and ran the hindcasts. The bred vector method is a technique used to generate perturbations that highlight the most unstable directions in the model's state space. By applying this method, we can create a diverse set of initial conditions that reflect the inherent uncertainties in the atmospheric state. This approach helps in generating a reasonably calibrated ensemble of hindcasts, which is crucial for understanding the range of possible future states of the atmosphere. 

We generated HENS as a dataset comprising 12 meteorological fields on a grid of size 721 × 1440 (latitude and longitude indices), corresponding to a 0.25-degree resolution (see Table \ref{table:variables}) with a temporal resolution of 6-hour intervals \cite{Mahesh2024hugeensemblespartii}.

\begin{table}[h!]
\caption{\textbf{The variables saved in the HENS output.}}
\centering
\begin{tabular}{l l r}
\toprule
\textbf{Variable}          & \textbf{Units}            & \textbf{Description}                    \\
\midrule
\texttt{t2m}               & K                        & 2m temperature                          \\
\texttt{d2m}               & K                        & 2m dew point temperature                \\ 
\texttt{tcwv}              & kg/m²                    & Total column water vapor                \\
\texttt{t850}              & K                        & Temperature at 850 hPa                  \\
\texttt{z500}              & m²/s²                    & Geopotential at 500 hPa                 \\
\texttt{z300}              & m²/s²                    & Geopotential at 300 hPa                 \\
\texttt{msl}               & Pa                       & Mean sea level pressure                 \\
\texttt{t500}              & K                        & Temperature at 500 hPa                  \\
\texttt{sp}                & Pa                       & Surface pressure                        \\
\texttt{ivt}               & kg/(m·s)                 & Integrated vapor transport              \\
\texttt{heat\_index}       & K                        & Heat index                              \\
\texttt{wind\_speed10m}       & m/s                      & 10m wind speed                          \\
\bottomrule
\end{tabular}
\label{table:variables}
\end{table}

\subsection{Observational records of tropical cyclones from summer 2023}
\label{ssec:IBTrACS}

The observational record for the tropical cyclones that occurred during Summer 2023 (Figure~\ref{fig:TC_Map}) is obtained from the International Best Track Archive for Climate Stewardship (IBTrACS) archive
\cite{Gahtana2024,Knapp2010}.  The International Best Track Archive for Climate Stewardship (IBTrACS) project is a comprehensive global collection of tropical cyclone data that combines recent and historical information from multiple agencies. Developed in collaboration with the World Meteorological Organization (WMO) and other organizations, IBTrACS has created a unified, publicly available dataset by collecting historical tropical cyclone data from Regional Specialized Meteorological Centers (RSMCs) and other agencies, merging the datasets, and disseminating them in formats used by the tropical cyclone community. IBTrACS provides location and intensity data for global tropical cyclones, spanning from the 1840s to the present, with data typically provided at 3-hour intervals. The dataset includes parameters such as position, intensity, and other related information, with some agencies providing additional data like radius of maximum winds and environmental pressure. By providing a unified dataset with summary statistics and original intensities from multiple agencies, IBTrACS enables a more comprehensive understanding of tropical cyclone climatology and uncertainty in intensity records.

\subsection{Reanalyses of meteorological states accompanying tropical cyclones}
\label{ssec:ERA5}

As a proxy for observations, we use the ERA5 global reanalysis prepared by the European Centre for Medium-Range Forecasting (ECMWF) \cite{Hersbach2020}.  This reanalysis provides all the data required to identify and track the evolution of hurricanes and tropical cyclones.  ERA5 is generated at hourly frequencies on a rectilinear latitude/longitude grid with a spatial resolution of 0.25${}^\circ$.   ERA5 is based upon a recent but earlier cycle (CY41R2) \cite{ECMWF79698} of the ECMWF Integrated Forecast System (IFS) \cite{ECMWF}.  As HENS is trained against ERA5 (with the period of interest withheld), the calibration of the TC detection and tracking framework we employ (section~\ref{ssec:TempestExtremes}) against the IBTrACS dataset (section~\ref{ssec:IBTrACS}) is directly transferable and applicable to HENS.

\subsection{TempestExtremes and Criteria Adjustments to TC Detection}
\label{ssec:TempestExtremes}

FourCastNet was originally trained to forecast meteorological variables and does not directly prognose the trajectory of TCs. Therefore, tracking TCs in HENS requires interpreting FourCastNet's output using a framework designed to identify and track the evolution of TCs. To this end, we will use TempestExtremes \cite{Ullrich_2017,Ullrich_2021}, a system developed to analyze extreme weather incidents by leveraging high-resolution weather hindcasts. This framework enables us to accurately monitor the formation and trajectory of TCs using the relevant meteorological variables generated by FourCastNet and contained in the HENS output.

The standard version of TempestExtremes \cite{Ullrich_2021} contains a generally applicable TC tracking system. The framework tracks TCs based on specific thresholds for latitude, \texttt{wind\_speed10m} (10-meter wind speed), \texttt{msl} (mean sea level pressure), and geopotential height, (\texttt{z300} - \texttt{z500}).  While generally effective, these thresholds required some adjustments to better match the tracks of actual TCs as recorded by IBTrACS (section~\ref{ssec:IBTrACS}) from the summer of 2023. The adjustments were generated by comparing the TCs detected by the TempestExtremes framework applied to the ERA5 dataset (section~\ref{ssec:ERA5})  against the actual TCs in the IBTrACS dataset.

To achieve this, we used an iterative process to maximize the hit rate and minimize the false alarm and false positive rates, i.e., maximizing (minimizing) the frequency of matches (mismatches)  between the TCs from IBTrACS and TempestExtremes, by making sequential adjustments to the thresholds employed in TempestExtremes.  At the conclusion of this optimization process, we find that the tracks of the tropical cyclones observed in summer 2023 can best be defined as a sequence of grid points in ERA5 satisfying the following criteria:
\begin{enumerate}
    \item \textbf{Position and Speed Criteria}
    \begin{itemize}
        \item The sequence must contain at least six distinct grid points drawn from anywhere along the track that are located between 5°S and 36.25°N.  
        \item The sequence must also contain at least six points drawn from anywhere along the track with a \texttt{wind\_speed10m} greater than 10 m/s.  
    \end{itemize}
    
    \item \textbf{Closed Contour Criteria}
    \begin{itemize}
        \item Starting from each grid point in the sequence,  the surface pressure \texttt{msl} must increase by $\ge$170 Pa over a 5.5° great-circle distance (GCD).  
        \item The difference in geopotential height (\texttt{z300} - \texttt{z500}) between the 300 hPa and 500 hPa surfaces must decrease by  $\ge$58.8 m²/s² over a 6.5° GCD originating from one or more grid points in the sequence, with the maximum value of this field occurring within a 1° GCD from these grid midpoints.  
    \end{itemize}
\end{enumerate}
  
The speed and localization criteria help reduce the number of falsely detected TCs, while the two closed contour criterion helped exclude cyclones that did not include a coherent upper-level warm core, which is a key meteorological feature characteristic of TCs. 

With these adjusted thresholds, we ran the TempestExtremes TC tracking framework on the ERA5 dataset.  In Table \ref{tab:hit_rate_false_alarm}, we catalog the performance metrics of the framework relative to the actual named tropical cyclones observed during summer 2023. We have improved the hit rate by 18\% and reduced the false alarm rate by 4\% relative to the default settings of the framework's thresholds. The formulae for these two metrics are given by:
\begin{eqnarray*}
\text{HR} &=& \frac{\text{Number of TCs identified with TE that were actually recorded during the summer}}{\text{Total number of TCs actually recorded over the summer (by agency)}} \\
\text{FAR} &=& \frac{\text{Number of TCs identified with TE that were not recorded during the summer}}{\text{Total number of potential TCs predicted with TE}} 
\end{eqnarray*}
These results show that the adjusted Tempest Extremes framework applied to ERA reanalysis correctly identifies tropical cyclones observed in the IBTrACS data set approximately 90\% of the time.  

We now apply this adjusted framework to detect tropical cyclones in the HENS ensemble.  At the initialization time $t_0$ for each 15-day hindcast, the initial conditions for the ensemble represent small perturbations relative to the coincident ERA reanalysis.  Since the Tempest Extremes framework exhibits good performance on ERA, and since the initial conditions of the hindcast approximate the coterminous reanalysis, we anticipate that the framework will identify tropical cyclones in HENS collocated with actual cyclones.  The validity of this hypothesis is demonstrated below.

\begin{table}[h]
\caption{\textbf{Performance evaluation of the TempestExtremes TC Tracking framework with different thresholds on the ERA dataset for Summer 2023 TCs.}}
\centering
\begin{tabular}{l r r}
\toprule
\textbf{Characterization} & \textbf{Hit Rate (HR)} & \textbf{False Alarm Rate (FAR)} \\
\midrule
Default thresholds \cite{Ullrich_2021} & 0.735 & 0.142 \\
Adjusted thresholds & 0.918 & 0.1 \\
\bottomrule
\end{tabular}
\label{tab:hit_rate_false_alarm}
\end{table}

\subsection{Tropical Cyclone Matching Algorithm}
\label{ssec:MatchingAlgorithm}
Given that we are able to accurately track TCs in ERA5, our next step is to match the tracks of actual TCs with tracks of TCs generated by HENS and identified using TempestExtremes operating under the adjusted thresholds. Since HENS is a multi-petabyte dataset, the process of track matching must be automated. The track-matching algorithm we have developed to perform this task consists of the following three steps as depicted in Figure~\ref{fig:TCMatchSchematic}:

\begin{enumerate}
    \item \textbf{Calculating the minimum distance between two TC tracks}: The first step is to define a distance metric between two TC tracks. The initial assumption is that if two TCs are not in the same tropical basin or if their dates of appearance are at least 15 days apart, they cannot be considered matches. We will label this set of conditions as configuration $(A)$. In configuration $(A)$, the distance between both TCs is set to $ +\infty $. In the complementary configuration $(\overline{A})$, we will define the distance between the two TCs using the Dynamic Time Warping (\textbf{DTW}) method combined with the Haversine, or great circle, shortest distance between two points on the Earth's surface. \\*[12pt]
    
    The DTW method measures the similarity between two sequences by aligning their temporal progression (Figure~\ref{fig:DTW}). This alignment is necessary because of variations between the detected and matching observed TC tracks; for example, sometimes the detected TC can start before or end after the matching observed TC or vice versa. The combination of the DTW algorithm applied to Haversine distances helps ensure more accurate matching of the TCs emulated by HENS with the actual ones. \\*[12pt]

    Let \( X = \{x_1, x_2, \ldots, x_N\} \) and \( Y = \{y_1, y_2, \ldots, y_M\} \) be two sequences of length \( N \) and \( M \), respectively. DTW can be formulated as follows:

    \[
    \text{DTW}_q(X, Y) =
        \min_{\pi \in \mathcal{A}(X, Y)}
            \left( \sum_{(i, j) \in \pi} d(x_i, y_j)^q \right)^{\frac{1}{q}}
    \]

    where
    $\mathcal{A}(X, Y)$ is the set of all possible alignments (warping paths) between sequences $X$ and $Y$, $d(x_i, y_j)$ is the distance function between elements $x_i$ and $y_j$ in the sequences, and $q$ is a parameter that determines the norm used to compute the cumulative distance. In our implementation, we set $d(\cdot,\cdot )$ to Haversine distances and set $q=1$. For a warping path $\pi$ to be considered valid, it must satisfy the following conditions:

    \begin{itemize}
        \item \textbf{Boundary Conditions}: The path must start at the beginning of both sequences, i.e., \( (1, 1) \), and end at the end of both sequences, i.e., \( (N, M) \).
        \item \textbf{Monotonicity}: The path must be monotonically non-decreasing in both dimensions, meaning that if \( (i_1, j_1) \) and \( (i_2, j_2) \) are two points in the path, then \( i_1 \leq i_2 \) and \( j_1 \leq j_2 \).
        \item \textbf{Continuity}: The path must consist of contiguous points such that each step in the path moves to either the next point in \( X \), the next point in \( Y \), or both.  Hence for each point \( (i_1, j_1) \) in a path, the subsequent point \( (i_2, j_2) \) may take on values of \( (i_1+1, j_1) \), \( (i_1, j_1+1) \), or \( (i_1+1, j_1+1) \).
    \end{itemize}

    \item \textbf{Applying perfect matching procedures to these minimum distances}: Having constructed the distance matrix between the emulated TCs and the actual ones, the next step is to use this matrix as input to a perfect matching procedure, such as the Hungarian procedure \cite{Wikipedia2025a} or the Stable Marriage procedure \cite{wikipedia2025b}. The Hungarian procedure finds the optimal one-to-one matching between the emulated and actual TCs that minimizes the total cost denominated in distances. Alternatively, the Stable Marriage procedure optimizes the one-to-one matching by requiring that the matches are ``stable.'' A stable matching is one in which no pair of elements would be better matched with each other rather than with their current matches, based on their ``preference lists.'' These preference lists are constructed according to the distances between the elements. \\*[8pt]
    
    Following the chosen perfect matching procedure, the initial distance matrix will need preprocessing. Since we are using the stable marriage procedure, we must ensure the matrix is square. If not, we introduce fictitious TCs that do not influence the matching process, and we also handle any ties.

    \item \textbf{Filtering on minimum distances between matching TCs at lead time~0}: At a lead time of~0, the state of each HENS ensemble member is nearly identical to its initial conditions from ERA5. The positions of the real TCs are unambiguously determined by these initial conditions, which closely reflect the actual meteorological conditions on the day in question. Therefore, since the TE framework has been calibrated to identify real TCs using these conditions, the positions of the emulated TCs should closely match the actual TCs at a lead time of~0. Analysis of the results of the TempestExtremes TC tracking framework at lead time~0 indicates that the initial distance between the actual TC and a good match should be less than \textbf{723 km}. We use this information to filter out poorly matching TCs, which generally correspond to matches between an actual TC and a false positive TC. This approach helps to significantly reduce the number of false positive TCs in the matches, although it may not eliminate them entirely.
\end{enumerate}

Before applying this matching algorithm to HENS, we tested it on ERA5 to check how it performed on identifying which TCs that have been hindcast using ERA5 correspond with actual TCs in the IBTrACS data set for summer~2023. We verified that we were able to perfectly match all correctly detected TCs.

Figure~\ref{fig:spread_error} shows that HENS is close to well-calibrated for TC track prediction: the ensemble spread grows with lead time in step with the ensemble-mean position error, from roughly 40–50 km at initialization to 500–550 km at 7 days. The spread–error ratio remains near 0.9 throughout the forecast (Figure~\ref{fig:spread_error}b), indicating only mild underdispersion, and its 95\% confidence interval includes the well-calibrated value of 1 at most lead times. The spread-error ratio of TC track forecasts has been used to assess the skill of operational ensemble forecast systems \cite{Hamill2013}, and this degree of calibration is comparable to that of operational ensemble prediction systems. This spread-error ratio  indicates that the track dispersion of HENS provides a meaningful estimate of forecast uncertainty. It ensures that HENS is a suitably reliable dataset with which to calculate the exponent of the physical diffusion process.  An important caveat is that this verification uses a single summer of storms (Table~A1), since HENS is only available over 1 summer due to its large size (3 petabytes).  This leaves substantial sampling uncertainty (shaded regions in Figure~\ref{fig:spread_error}) and can be addressed in future large ensemble experiments. 

\section{Results} 
\label{sec:Results}
\begin{figure}
\centering
\begin{subfigure}{.4\linewidth}
  \includegraphics[width=\linewidth]{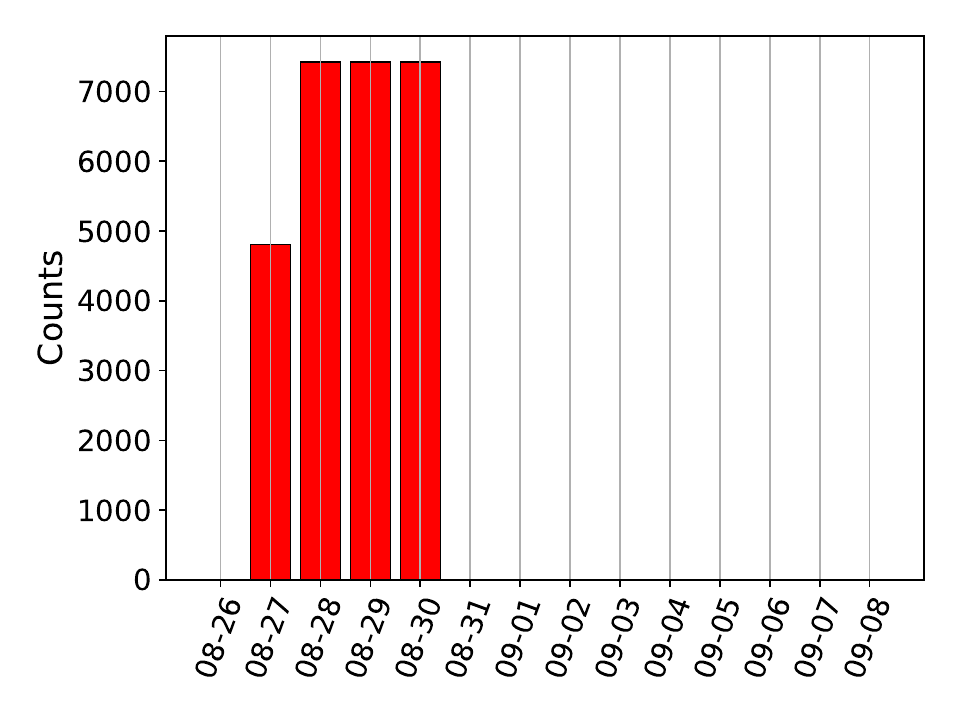}
  \caption{Idalia}
\end{subfigure}\hspace{1em}
\begin{subfigure}{.4\linewidth}
  \includegraphics[width=\linewidth]{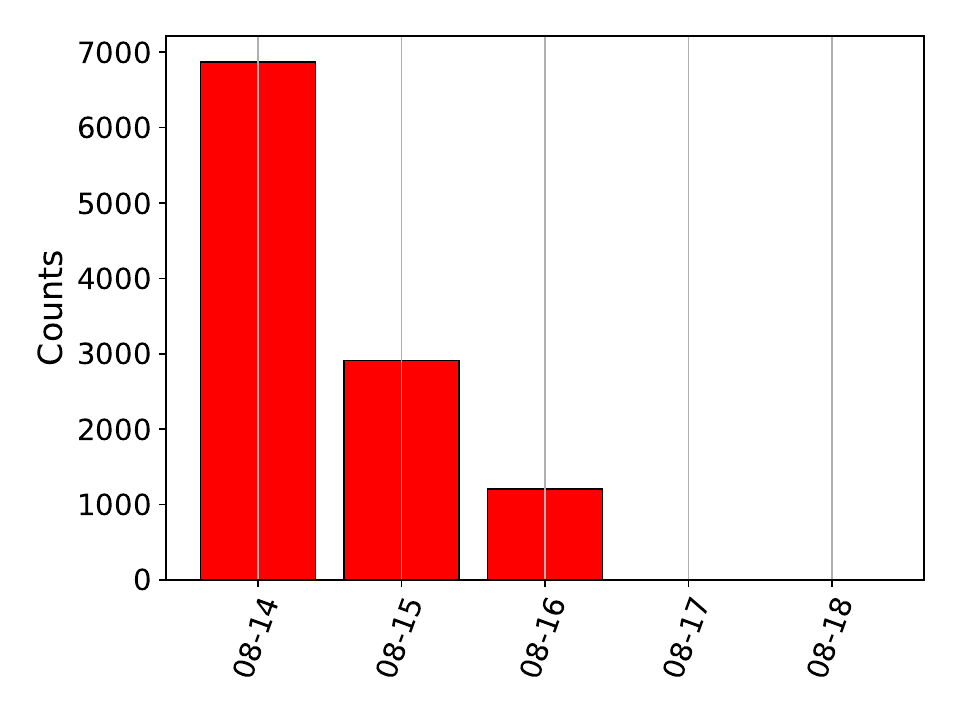}
  \caption{Greg}
\end{subfigure}

\medskip
\begin{subfigure}{.4\linewidth}
  \includegraphics[width=\linewidth]{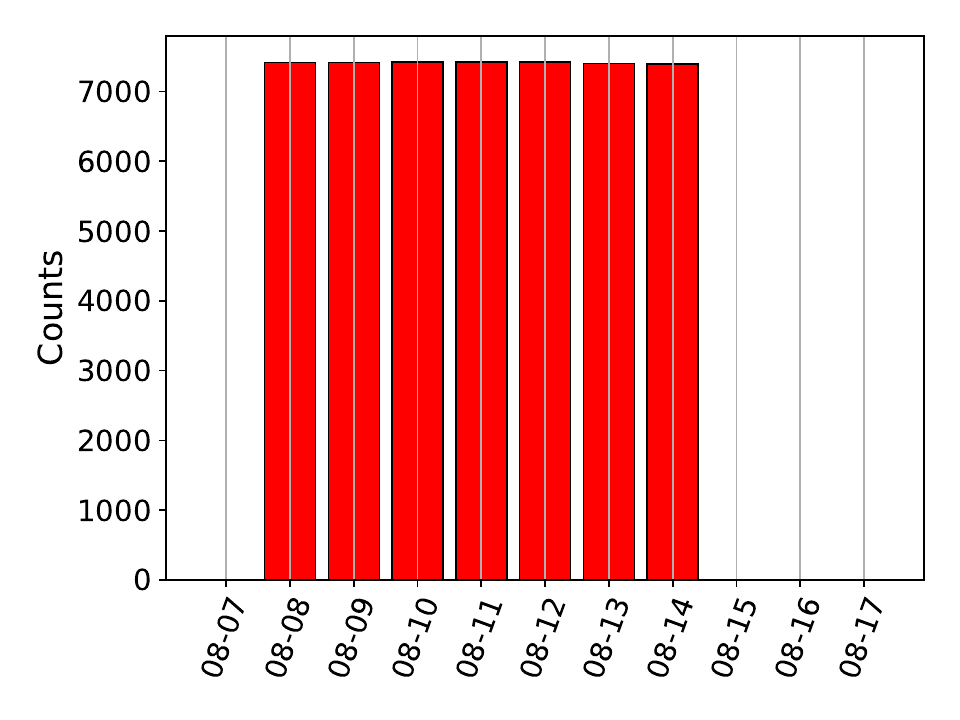}
  \caption{Lan}
\end{subfigure}\hspace{1em}
\begin{subfigure}{.4\linewidth}
  \includegraphics[width=\linewidth]{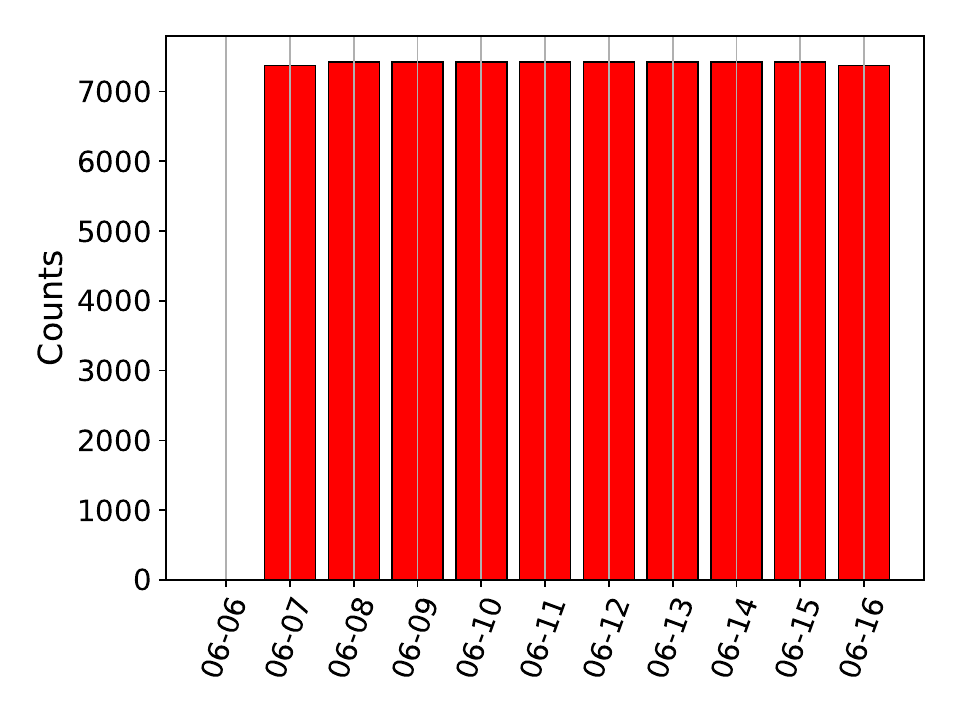}
  \caption{Biparjoy}
\end{subfigure}
\caption{\textbf{The numbers of matched replicates for each of the four representative TCs ($\S$\ref{ssec:TCsStudied}) versus the  initialization dates for the 15-day HENS hindcasts from summer of 2023.} The ensembles of 15-day hindcasts are identified by the dates shown on the x-axis when each is initialized. The numbers are plotted between the formation and dissipation of the corresponding representative TC.  The maximum possible number of matches equals 7424, the number of ensemble members in each hindcast.}
\label{fig:distribution_tc_hens}
\end{figure}

\subsection{Introduction: List of TCs Studied}
\label{ssec:TCsStudied}
We focus on tropical cyclones (TCs) from the summer of 2023  that started or ended between June 1 and August 31. See Table~\ref{tab:TC_2023} for the list of the 33 observed TCs that occurred during this period and Figure~\ref{fig:TC_Map} for a map of the corresponding tracks.

After running the TempestExtremes TC tracking framework ($\S$\ref{ssec:TempestExtremes}) on HENS, we applied the matching algorithm ($\S$\ref{ssec:MatchingAlgorithm}) to associate the emulated TCs with the 33 actual TCs. To illustrate the properties of the emulated TCs, we will focus on a specific observed TC from each of four tropical cyclone basins and will characterize the spatio-temporal distributions and ensemble spreads of TCs produced by HENS associated with those real TCs. The basins are the North Atlantic, Eastern Pacific, Western Pacific, and North Indian Ocean. The illustrative TCs  include Idalia (North Atlantic, from August 26 to August 31), Greg (Eastern Pacific, from August 14 to August 18), Lan (Western Pacific, from August 27 to September 6), and Biparjoy (North Indian Ocean, from June 6 to June 19). Figure \ref{fig:distribution_tc_hens} shows how many HENS ensemble members contain emulated TCs that are closely associated (matched) with the actual illustrative TCs as a function of the initialization dates for the hindcasts. For example, Biparjoy is replicated in over 7000 ensemble members drawn from 15-day hindcasts initialized on 10 consecutive days in June 2023.  

\subsection{Computation of Exponents Along TC Paths: Histograms and CDFs}
\label{ssec:Exponents}
The fact that each actual TC is replicated by several thousand closely matched counterfactual TCs in HENS makes it possible to reliably compute the diffusion of these counterfactual tracks as a function of time.  In contrast to earlier studies \cite<e.g.,>{Meuel:hal-00708481} that characterized the diffusive properties of TCs using several hundred independent TCs in observations, here we can fit the diffusion power laws using several thousand counterfactual copies of a single observed TC generated by learnt AI weather dynamics emulation. 

Using the huge ensemble of counterfactual track data, we  calculate the mean square displacement $\text{MSD}_n'(t)$ as a function of time $t$ across the ensembles of replicates matched to each observed TC numbered $n = 1 
\ldots 33$ (Table~\ref{tab:TC_2023}). The form of $\text{MSD}_n'(t)$ we adopt to measure the mean-square separations among the positions \( \vec{X}(t) \) of the matching HENS ensemble members is
\begin{equation}
  \text{MSD}_n'(t) =  \frac{\left\llangle \Big(\vec{X}_j(t + t_0) - \big\llangle \vec{X}_j(t + t_0)\big\rrangle \Big )^2 \right\rrangle}{\left\llangle \Big(\vec{X}_j(t_0) - \big\llangle \vec{X}_j( t_0)\big\rrangle \Big )^2 \right\rrangle} 
  \label{eq:MSDens}
\end{equation}   
where $\llangle \ldots \rrangle$ denotes the average over matching ensemble members indexed by $j$.  The $\text{MSD}'(t)$ quantifies the growth in the separations among the ensemble members relative to the ensemble-average TC track.  Time $t$ is measured relative to an initial time $t_0$ when a given named TC (\ref{app:TC_Tables}) first appears in HENS.   

\begin{figure}
\centering
\begin{subfigure}{.4\linewidth}
  \includegraphics[width=\linewidth]{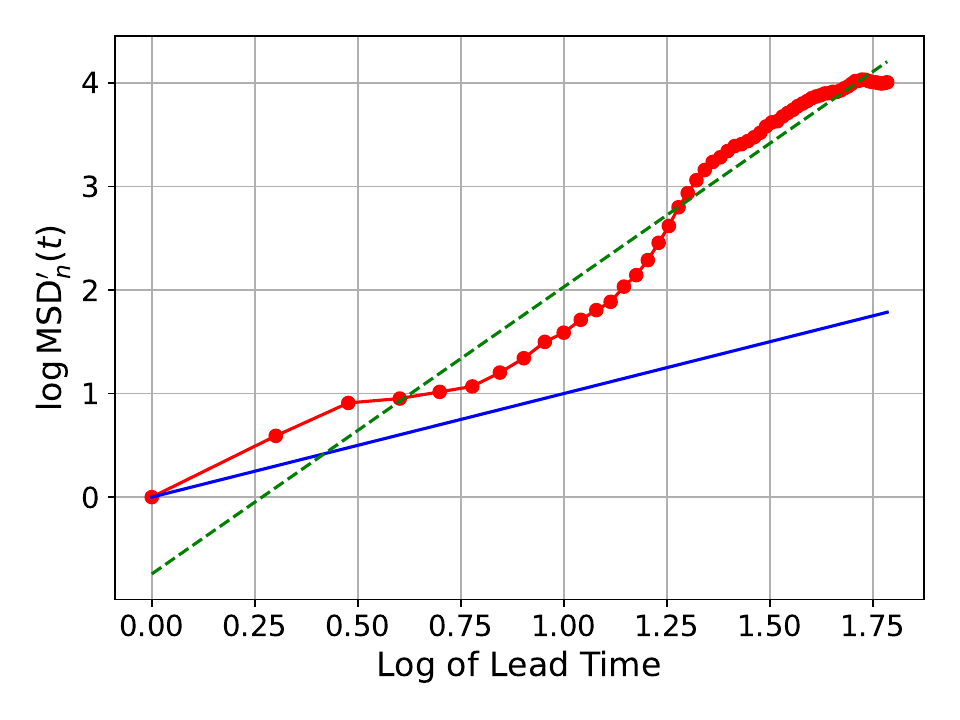}
  \caption{Idalia}
\end{subfigure}\hspace{1em}
\begin{subfigure}{.4\linewidth}
  \includegraphics[width=\linewidth]{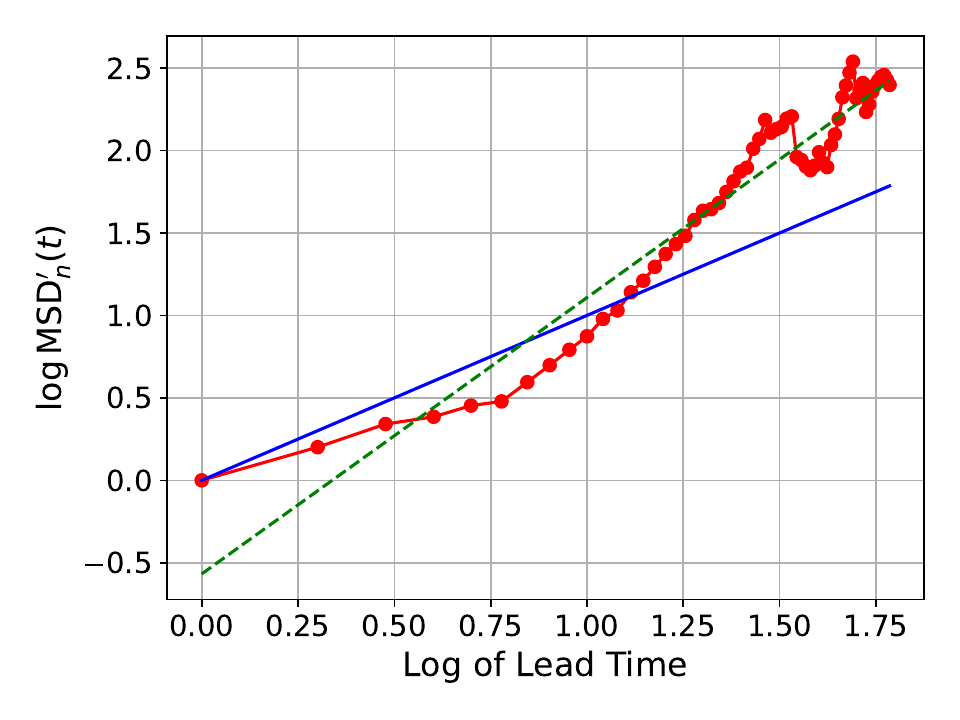}
  \caption{Greg}
\end{subfigure}

\medskip
\begin{subfigure}{.4\linewidth}
  \includegraphics[width=\linewidth]{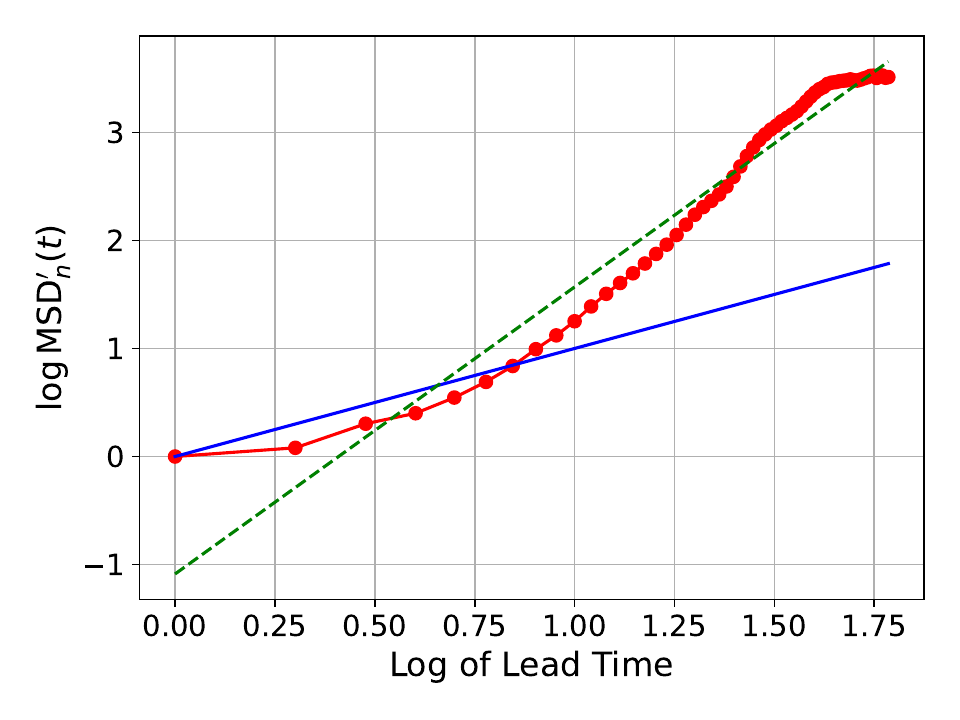}
  \caption{Lan}
\end{subfigure}\hspace{1em}
\begin{subfigure}{.4\linewidth}
  \includegraphics[width=\linewidth]{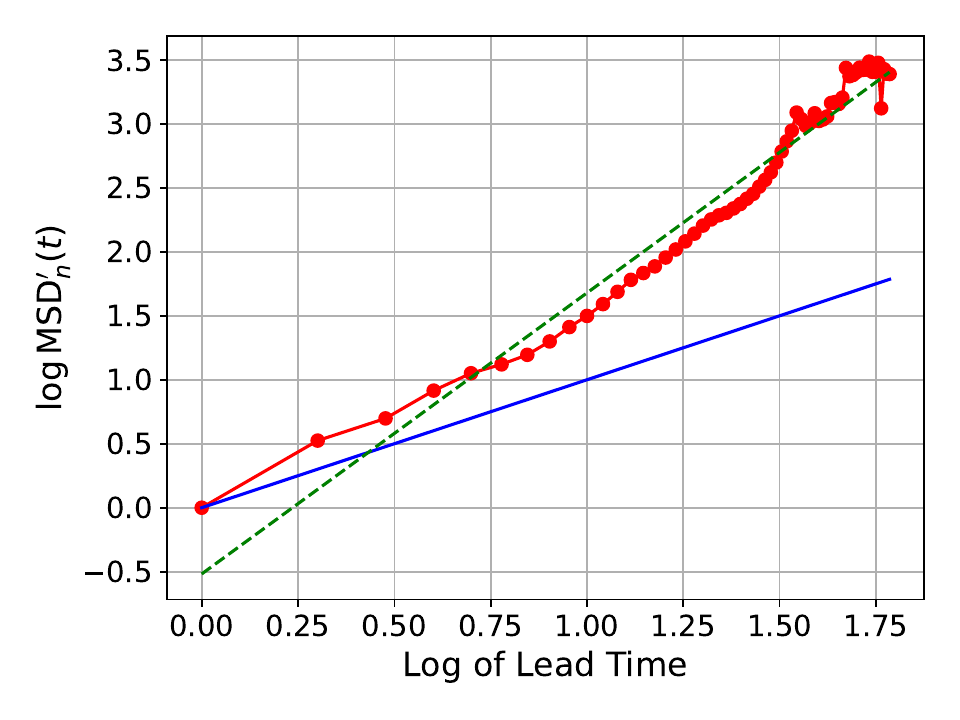}
  \caption{Biparjoy}
\end{subfigure}
\caption{\textbf{Log-log plots of the variance in centroid positions of the HENS TCs versus time $t$ since time of initialization $t_0$.} Each red point represents a 6-hourly calculation of $\textbf{MSD}'_n(t)$ (Eq.~\ref{eq:MSDens}). Green dashed lines show least-squares fits to Eq.~\ref{eq:exponent_eprime}. Solid blue lines depict Eq.~\ref{eq:exponent_eprime} with $e' = 1$.  The x axis refers to the $\text{log}_{10}$ of the number of time steps, or 6-hour intervals, in the ML model rollout.}
\label{fig:measures_tc_hens}
\end{figure}

Following earlier studies of the (possibly anomalous) diffusive behavior of atmospheric dynamics, we hypothesize that  $\text{MSD}'(t)$ can be approximated as a power law as
\begin{equation}
   \text{MSD}'_n(t) \propto t^{e'_n}
\label{eq:exponent_eprime}
\end{equation}
For classical diffusion, theory dictates that $e'_n = 1$.
Log-log plots of $\textbf{MSD}'_n(t)$ (Eq.~\ref{eq:MSDens}) and its power-law approximation (Eq.~\ref{eq:exponent_eprime}) for the four representative TCs are shown in Figure~\ref{fig:measures_tc_hens}.  
For several of these TCs, in particular Biparjoy, a power law in time is indeed a reasonable approximation to the temporally evolving variance in positions across the matching HENS ensemble members. It is evident from this figure that the exponents $e'_n$ that best fit the data appreciably exceed 1.  This exceedance implies that the HENS ensembles of replicate TCs are consistently exhibiting super-diffusive behavior for all four representative TCs.  

We can also test whether individual replicates are exhibiting super-diffusive behavior. Following \cite{Meuel:hal-00708481}, the MSD for individual ensemble members indexed by $j$ can be computed by
\begin{equation}
    \text{MSD}_j(t) = \left\langle \Big(\vec{X}_j(t + t') - \vec{X}_j(t')\Big)^2 \right\rangle
    \label{eq:MSD}
\end{equation}
where $\langle \ldots \rangle$ denotes an average over $t' \ge t_0$ and $t_0$ is again the time of emergence of the TC.   
In keeping with prior studies of individual TCs \cite<e.g.,>{Blender1997}, we find in general that the MSD can be accurately approximated by a power law in time, i.e., 
\begin{equation}
   \text{MSD}_j(t) \propto t^{e_j} \label{eq:exponent_e}
\end{equation}
This relates the squared displacement to a power law in $t$ with an exponent \( e_j \) characterizing the diffusion behavior of this particular ensemble member.  

\begin{figure}
\centering
\begin{subfigure}{.4\linewidth}
  \includegraphics[width=\linewidth]{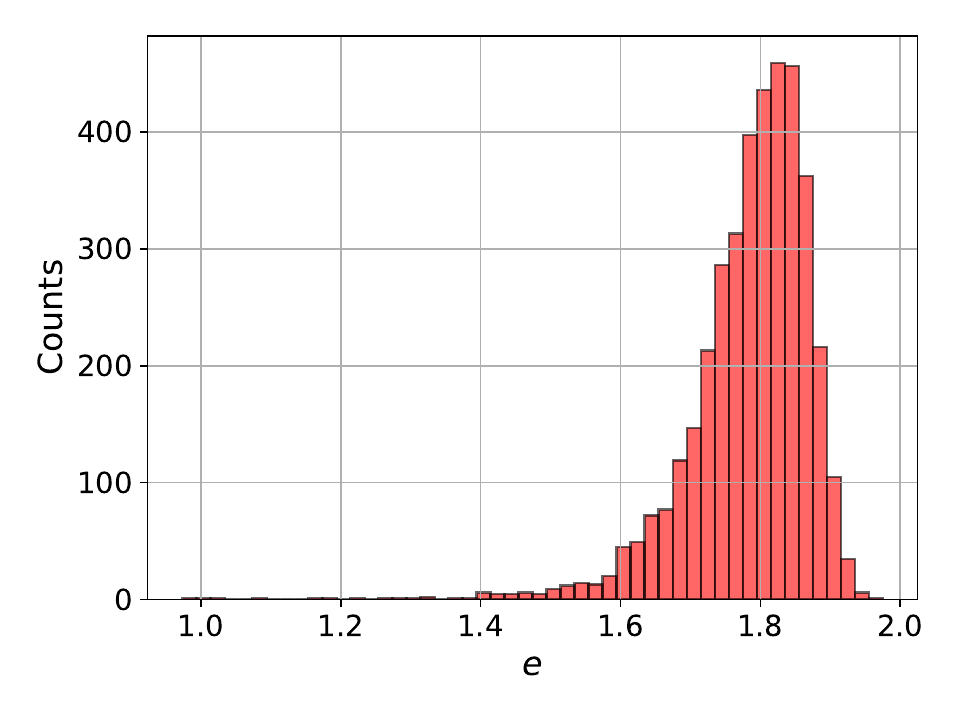}
  \caption{Idalia (H)}
\end{subfigure}\hspace{1em}
\begin{subfigure}{.4\linewidth}
  \includegraphics[width=\linewidth]{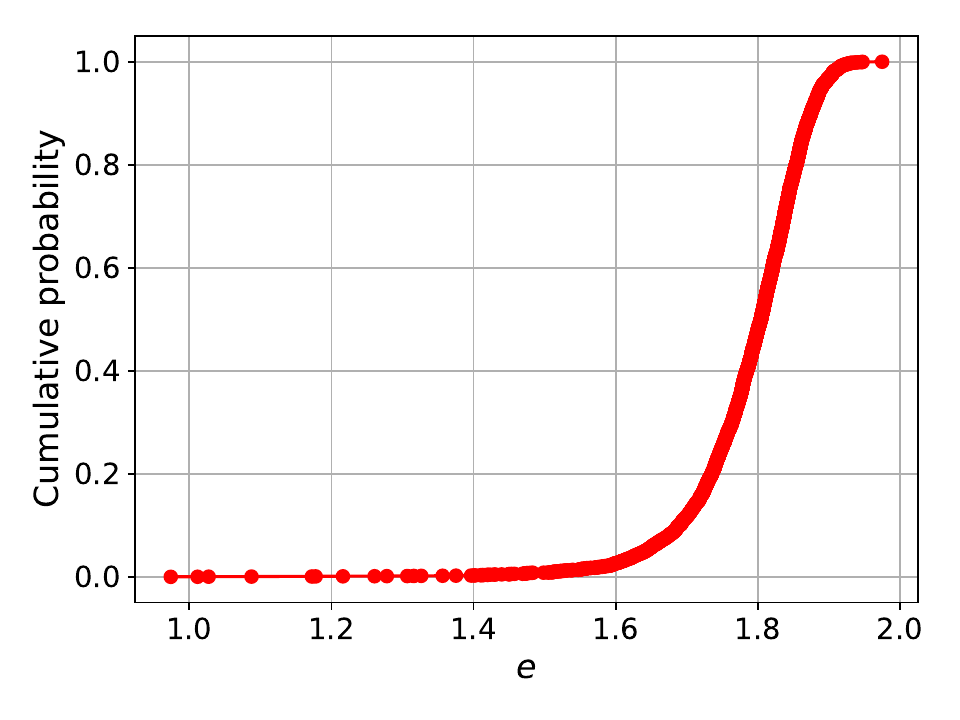}
  \caption{Idalia (C)}
\end{subfigure}

\medskip
\begin{subfigure}{.4\linewidth}
  \includegraphics[width=\linewidth]{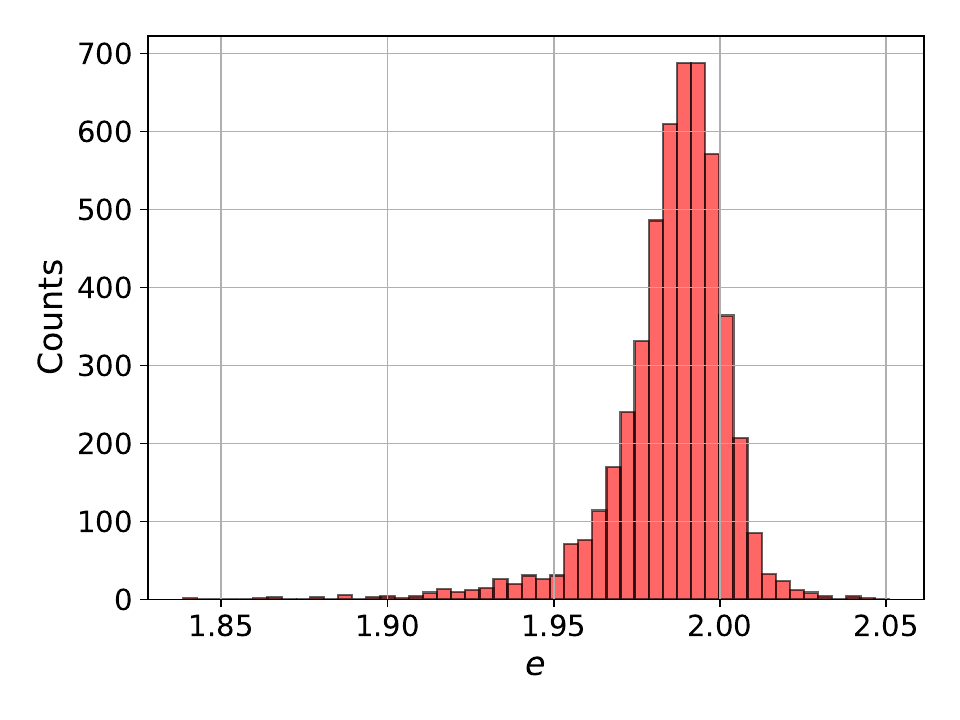}
  \caption{Greg (H)}
\end{subfigure}\hspace{1em}
\begin{subfigure}{.4\linewidth}
  \includegraphics[width=\linewidth]{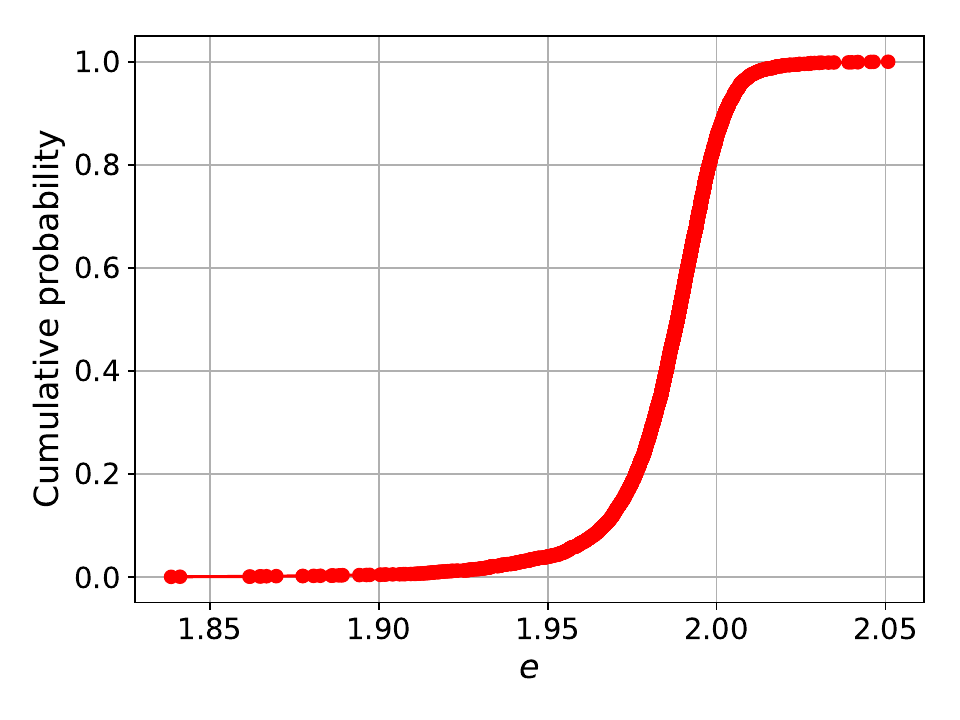}
  \caption{Greg (C)}
\end{subfigure}

\medskip
\begin{subfigure}{.4\linewidth}
  \includegraphics[width=\linewidth]{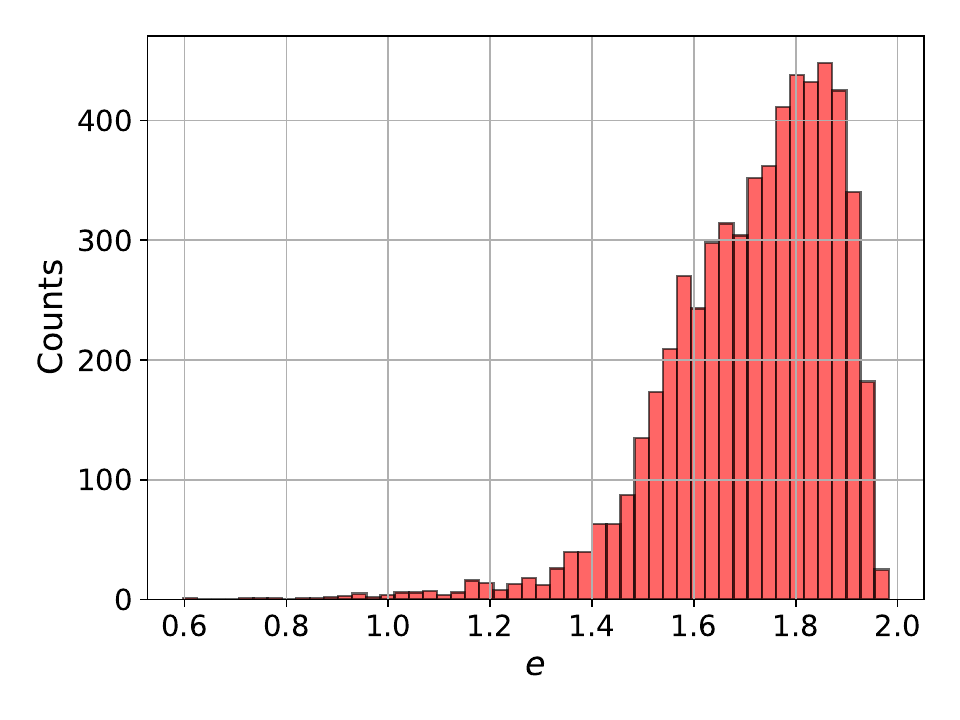}
  \caption{Lan (H)}
\end{subfigure}\hspace{1em}
\begin{subfigure}{.4\linewidth}
  \includegraphics[width=\linewidth]{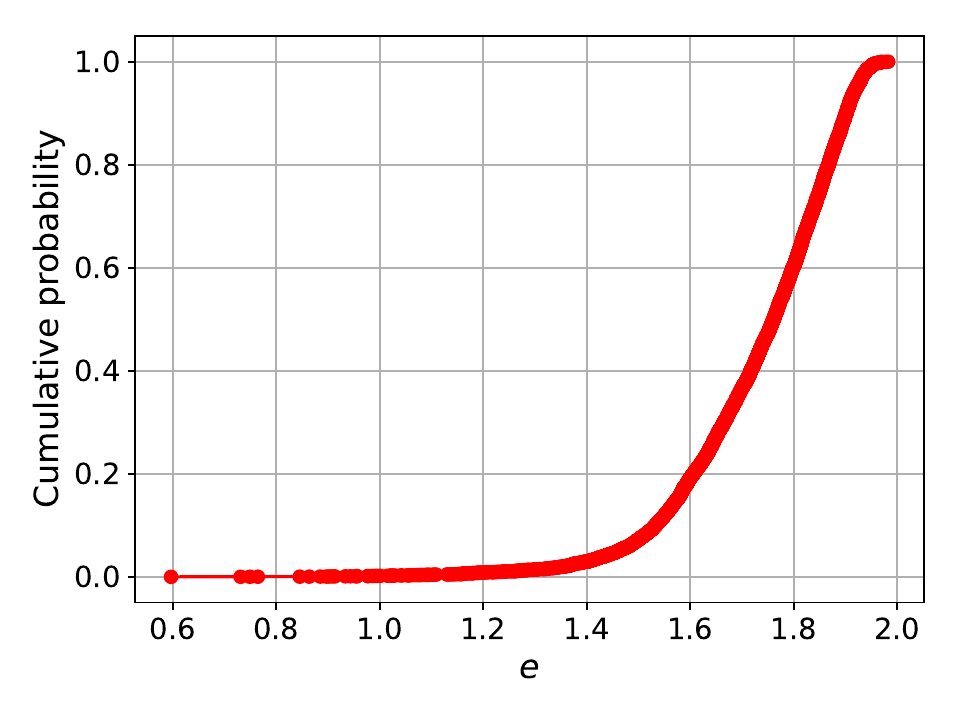}
  \caption{Lan (C)}
\end{subfigure}

\medskip
\begin{subfigure}{.4\linewidth}
  \includegraphics[width=\linewidth]{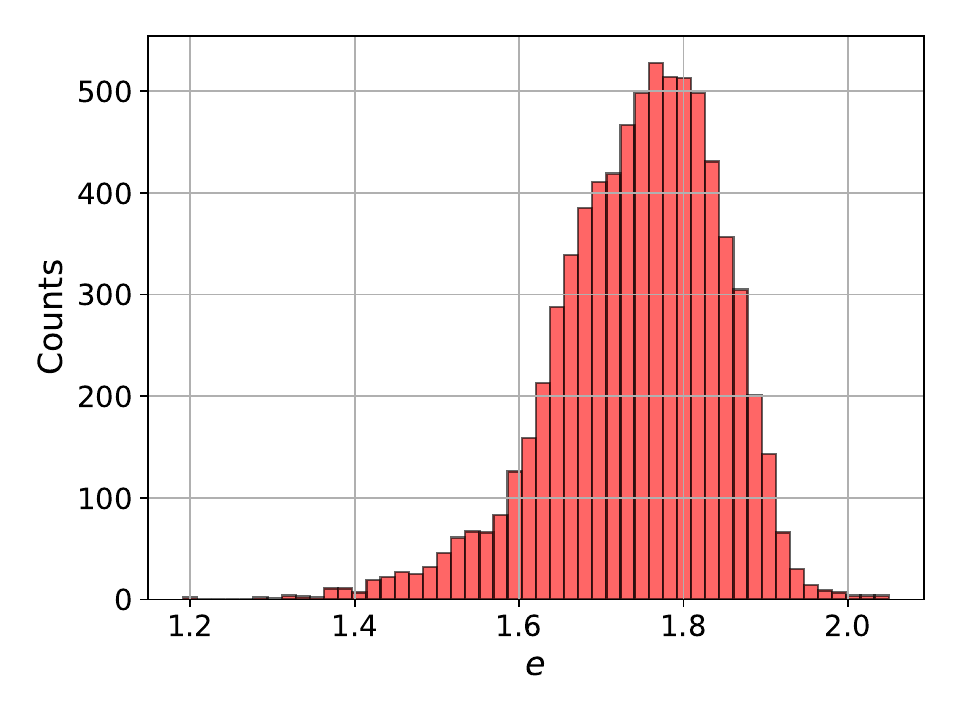}
  \caption{Biparjoy (H)}
\end{subfigure}\hspace{1em}
\begin{subfigure}{.4\linewidth}
  \includegraphics[width=\linewidth]{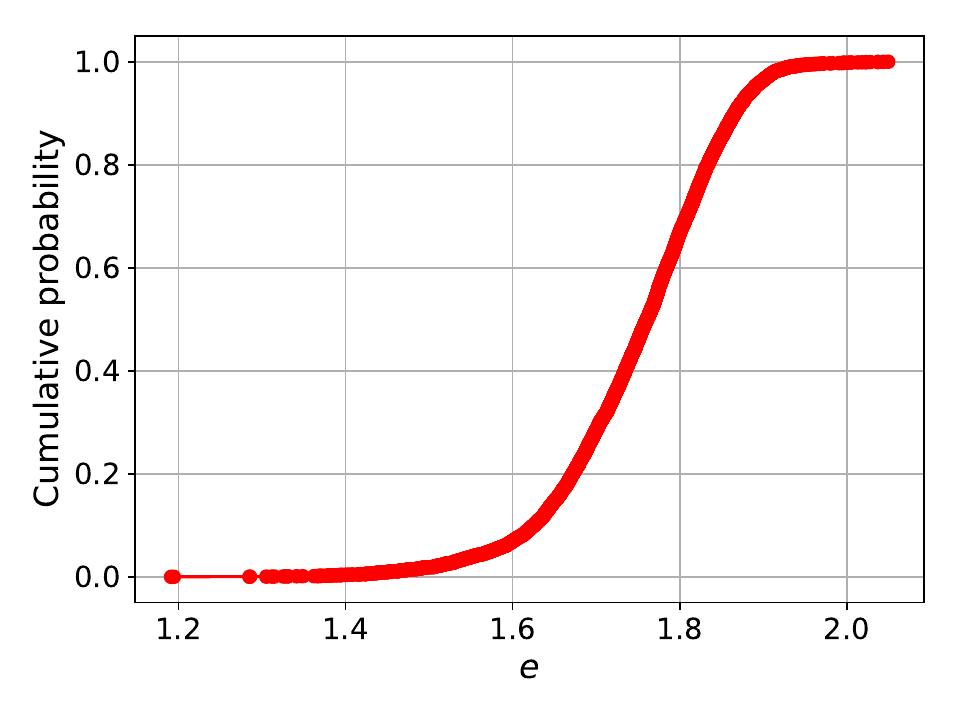}
  \caption{Biparjoy (C)}
\end{subfigure}

\caption{\textbf{Histograms (labeled ``H'') and  cumulative probability distributions (labeled ``C'') of the track diffusion exponent \(e\)  for each of the selected TCs ($\S$\ref{ssec:TCsStudied}).} The approximate power-law growth with time of the mean-square displacement of each TC from its initial position at its start time $t_0$ (Eq.~\ref{eq:MSD}) is quantified by $e$ (Eq.~\ref{eq:exponent_e}).  Note the different x axis scale used for each TC.}
\label{fig:tc_hens_combined}
\end{figure}

Histograms of the number of ensemble members with specific values of $e_j$ and cumulative distribution functions (CDFs) of $e_j$ for the four representative TCs are shown in Figure~\ref{fig:tc_hens_combined}.  Three features of these plots are evident from inspection.  First, the vast majority of individual ensemble members exhibit super-diffusive behavior with $e_j > 1$.  Second, very few ensemble members exceed the ballistic limit of $e_j = 2$. Third, the sigmoid shapes of the HENS CDFs qualitatively and quantitatively resemble the CDFs of exponents accumulated over hundreds of observed TCs \cite{Meuel:hal-00708481}.

These three features are a general property of the HENS emulations of the TCs observed during summer 2023.  The 25th, 50th, and 75th percentile values of exponents $e_j$ are tabulated in \ref{app:TC_Tables} for the 33 named TC systems.  In every case these values exceed 1 and are bounded above by 2.  

HENS enables calculating a distribution of $e_j$ across thousands of ensemble members for each individual storm (Figure~\ref{fig:tc_hens_combined}).  In contrast, observations provide only a single estimate of $e$ per storm, so characterizing the distribution requires aggregating across multiple storms.  This is the prevailing approach in the literature \cite{Meuel:hal-00708481}, although it masks the distinct distributions revealed from an ensemble of predictions for each storm (Figure~\ref{fig:tc_hens_combined}).  Figure~\ref{fig:ibtracs_msd_histogram} shows that $e$ in observations is similarly constrained between 1 and 2, with a noticeable skew toward the ballistic limit.  This super-diffusive skew is consistent with the per-storm HENS distributions (Figure~\ref{fig:distribution_tc_hens}).

\newpage

\section{Discussion}
\label{sec:Discussion}
In addition to purely scientific considerations, reliable predictions of tropical cyclone and hurricane tracks with sufficient lead time could substantially facilitate proactive disaster preparedness and societal risk mitigation.  The absolute errors in operational TC predictions measured in, e.g., nautical miles, for fixed lead times have steadily declined as numerical weather prediction (NWP) systems have advanced in the recent past \cite{Zhou_2020,Yu_2022}.  There is ongoing debate as to whether further absolute track errors will asymptotically approach a fixed lower bound depending on lead time, i.e., ``stall'' \cite{Landsea_2018} or continue to decrease gradually as NWP systems continue to improve \cite{Zhou_2020,Yu_2022}.  

In the context of this paper, the mean-square displacements (MSDs) analyzed in $\S$\ref{sec:Results} represent well-characterized estimates of uncertainty.  Two independent parameters govern the magnitudes of these uncertainties: the uncertainty $\textbf{MSD}(t_0)$ at the time $t_0$ of initiation at the TC in question and the exponent $e$ governing the anomalous diffusion of the TC in the background flow.  MSDs increase monotonically with both independent parameters, with linear dependence on  $\textbf{MSD}(t_0)$ and exponential dependence on $e$.  It is essentially self-evident that more accurate initial conditions will reduce the magnitudes of $\textbf{MSD}(t_0)$ and will thereby lower the uncertainty $\textbf{MSD}(t)$ at all subsequent lead times $t$.  However, in the absence of a theory or empirical relationship determining the dependence of $e$ on the properties of the TC and its surrounding synoptic conditions (and errors in their forecasts), it is not clear whether and how $e$ could be reduced through improvements in NWP systems. The authors suggest that systematic investigation of the local and large-scale meteorological factors governing $e$ could potentially be very useful for reducing TC track errors in the future.

\section{Conclusions}
\label{sec:Conclusions}

We show that huge ensembles  of ML-based hindcasts provide a novel statistical framework for analyzing the diffusive properties of TCs. HENS provides thousands of counterfactual recreations of each individual storm, confirming the anomalous super-diffusive behavior previously inferred from observational records. We demonstrate that the mean-squared displacement of TC tracks follows a power-law relationship with time, characterized by exponents $e$ that consistently indicate super-diffusion ($e > 1$) yet rarely surpass the ballistic limit $(e < 2)$. Crucially, the distribution of diffusion exponents can be constructed from the ensemble spread of a single storm, rather than from the aggregate of all historical observations.  This capability is uniquely enabled by the massive sample sizes that HENS provides. These results establish that ML-based weather emulators can be used to characterize the dynamics of TC propagation through the atmospheric flow. The computational efficiency of ML emulators makes them a powerful new tool in atmospheric science, enabling analyses that would be prohibitively expensive with conventional numerical models.

Our findings have important implications for the predictability of tropical cyclone tracks and landfall risks on medium-range time scales.  Two key components of TC track forecast uncertainty are the initial-condition error at the time of cyclone emergence and the model uncertainty that also governs subsequent track dispersion. While improvements in observational networks and data assimilation can systematically reduce the former, our findings suggest that advances in ensemble weather forecasts must also account for the super-diffusive nature of TC tracks. The diffusion exponent $e$ reflects an intrinsic property of the interaction between a given TC and its large-scale steering environment, and the super-diffusive nature of TCs drives a rapid expansion of potential landfall locations with increasing lead time. Future research should focus on identifying the specific meteorological drivers that govern variations in this exponent across storms, basins, and synoptic regimes. Such understanding is essential for moving beyond the current limits of NWP ensemble sizes and enhancing societal resilience to TC hazards.

\newpage
\appendix
\section{Tropical Cyclones Tables} 
\label{app:TC_Tables}
\begin{table}[ht]
\centering
\caption{Ensemble-Median Exponents for Tropical Cyclones Observed in Summer 2023.
For each TC, the medians of the exponents (Eq.~\ref{eq:exponent_e}) computed from each matching hindcast are collected in series.  The table reports the 25th, 50th, and 75th percentile values from this series. 
Dates are in DD/MM format.  The basins are North Atlantic (NA), North Indian (NI), Eastern Pacific~(EP), and WP (Western Pacific).}
\begin{tabular}{l l l l r r r}
\toprule
\textbf{TC index} & \textbf{TC name} & \textbf{Dates active} & \textbf{Basin} & $e^{ med}_{25\%}$ & $e^{ med}_{50\%}$ & $e^{ med}_{75\%}$ \\
\midrule
 1 & Mawar (Betty) & 19/05 – 03/06 & WP & 1.87 & 1.88 & 1.89 \\
 2 & Arlene & 01/06 – 03/06 & NA & 1.80 & 1.80 & 1.80 \\
 3 & Guchol (Chedeng) & 05/06 – 12/06 & WP & 1.74 & 1.86 & 1.94 \\
 4 & Biparjoy & 06/06 – 19/06 & NI & 1.76 & 1.79 & 1.92 \\
 5 & Bret & 19/06 – 24/06 & NA & 1.95 & 1.97 & 1.97 \\
 6 & Cindy & 22/06 – 26/06 & NA & 1.95 & 1.96 & 1.97 \\
 7 & Adrian & 27/06 – 02/07 & EP & 1.94 & 1.96 & 1.97 \\
 8 & Beatriz & 29/06 – 01/07 & EP & 1.93 & 1.93 & 1.93 \\
 9 & Calvin & 11/07 – 19/07 & EP & 1.97 & 1.98 & 1.99 \\
10 & Talim (Dodong) & 13/07 – 18/07 & WP & 1.90 & 1.90 & 1.93 \\
11 & Don & 14/07 – 24/07 & NA & 1.58 & 1.60 & 1.75 \\
12 & Doksuri (Egay) & 20/07 – 30/07 & WP & 1.89 & 1.91 & 1.93 \\
13 & Four-E & 21/07 – 22/07 & EP & 1.99 & 1.99 & 1.99 \\
14 & Khanun (Falcon) & 26/07 – 12/08 & WP & 1.59 & 1.73 & 1.84 \\
15 & BOB 03 & 31/07 – 03/08 & NI & 1.69 & 1.77 & 1.79 \\
16 & Dora & 31/07 – 21/08 & WP & 1.95 & 1.96 & 1.97 \\
17 & Eugene & 05/08 – 07/08 & EP & 1.90 & 1.99 & 2.00 \\
18 & Lan & 07/08 – 17/08 & WP & 1.79 & 1.84 & 1.85 \\
19 & Fernanda & 12/08 – 17/08 & EP & 1.96 & 1.98 & 1.98 \\
20 & Greg & 14/08 – 18/08 & EP & 1.96 & 1.97 & 1.98 \\
21 & Hilary & 16/08 – 21/08 & EP & 1.86 & 1.90 & 1.93 \\
22 & Gert & 19/08 – 04/09 & NA & 1.67 & 1.81 & 1.90 \\
23 & Emily & 20/08 – 21/08 & NA & 1.83 & 1.84 & 1.85 \\
24 & Franklin & 20/08 – 01/09 & NA & 1.64 & 1.66 & 1.68 \\
25 & Damrey & 21/08 – 29/08 & WP & 1.59 & 1.62 & 1.65 \\
26 & Harold & 21/08 – 23/08 & NA & 1.96 & 1.96 & 1.96 \\
27 & Saola (Goring) & 22/08 – 03/09 & WP & 1.59 & 1.77 & 1.81 \\
28 & Idalia & 26/08 – 31/08 & NA & 1.79 & 1.81 & 1.82 \\
29 & Haikui (Hanna) & 27/08 – 06/09 & WP & 1.85 & 1.86 & 1.88 \\
30 & Irwin & 27/08 – 29/08 & EP & 1.89 & 1.97 & 1.99 \\
31 & Kirogi & 29/08 – 04/09 & WP & 1.92 & 1.93 & 1.95 \\
32 & Jose & 29/08 – 02/09 & NA & 1.13 & 1.15 & 1.17 \\
33 & Katia & 31/08 – 04/09 & NA & 1.89 & 1.91 & 1.92 \\ 
\bottomrule
\end{tabular}
\label{tab:TC_2023}
\end{table}

\newpage
\section{Tropical Cyclone Track Matching Algorithm Schematic}
\begin{figure}[h!]
    \makebox[\textwidth][c]{%
        \includegraphics[width=1.3\textwidth]{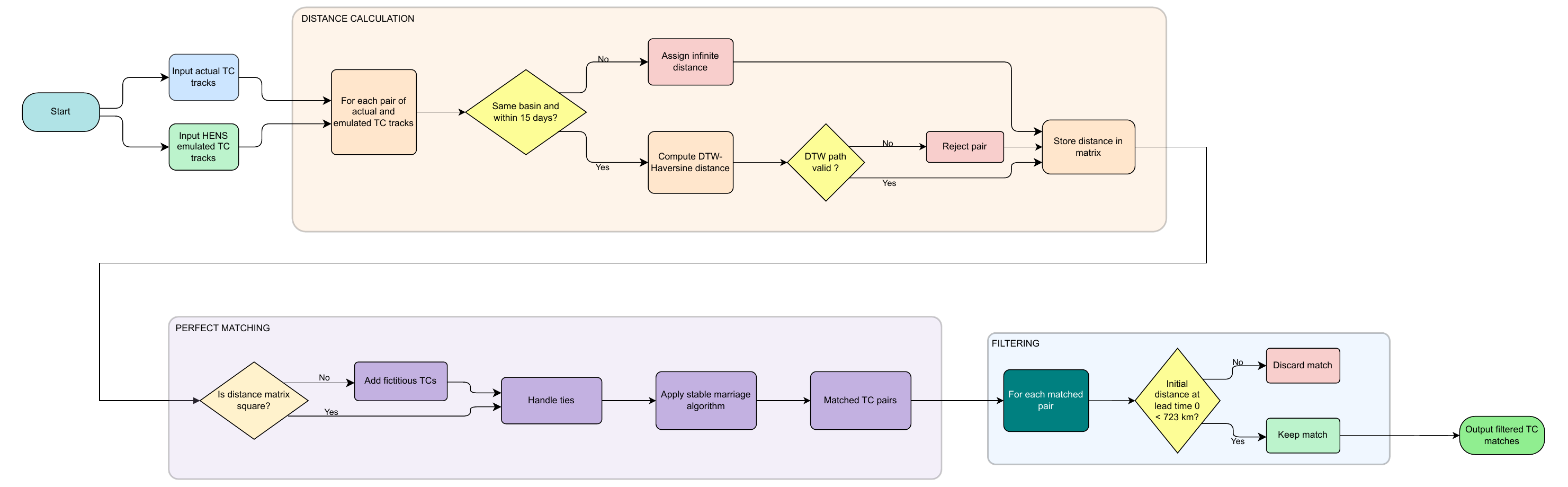}%
    }
\caption{Flowchart of the tropical cyclone matching algorithm, showing distance computation, filtering, tie handling, and the final matched TC pairs.} \label{fig:TCMatchSchematic}
\end{figure}

\begin{figure}[h!]
\centering
\includegraphics[width=0.90\textwidth]{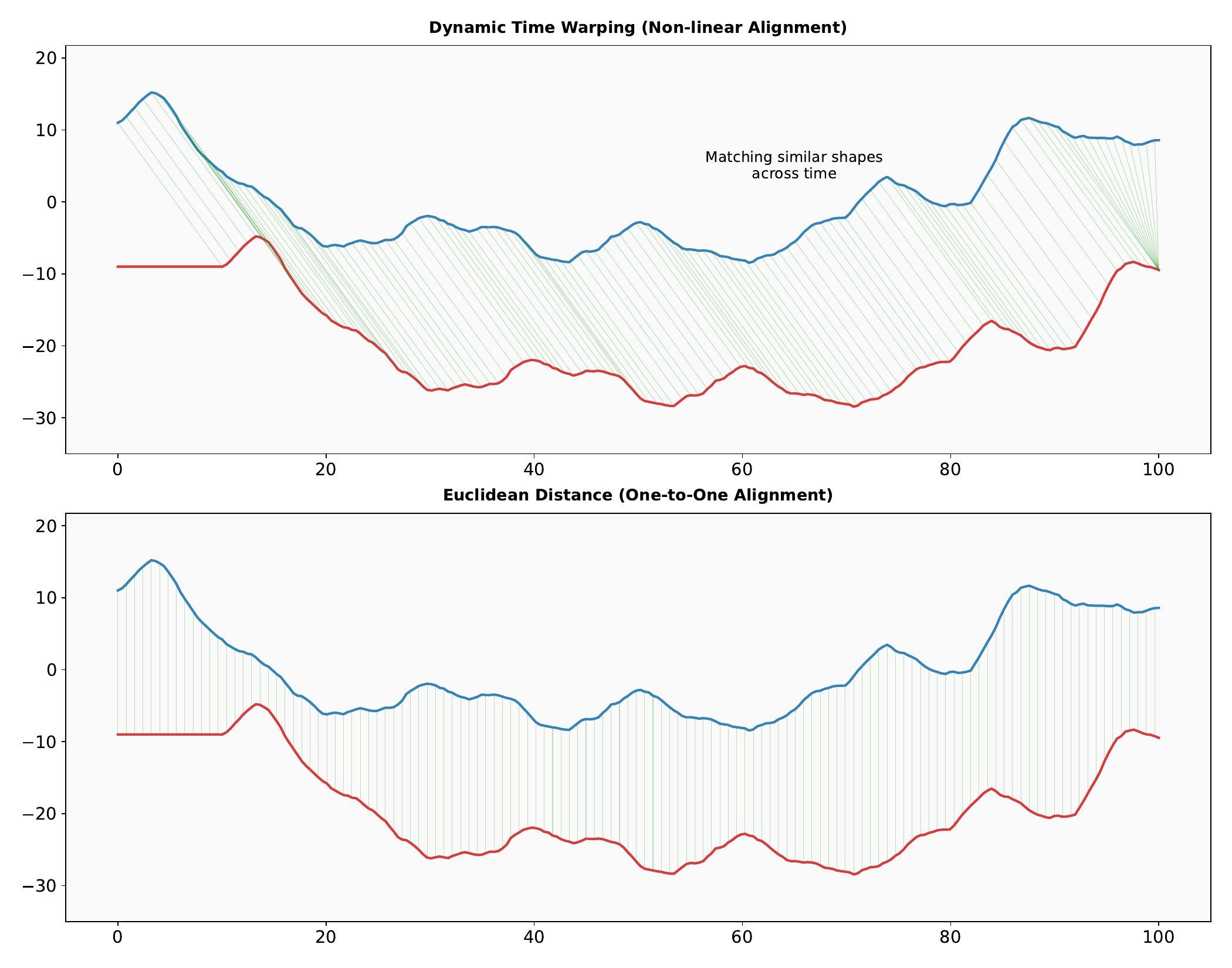}
\caption{Illustration of the difference between DTW distance and Euclidean distance (green lines indicate the alignment between points of time series A and B). Time series B corresponds to a time-shifted variant of time series A. Unlike Euclidean distance, DTW focuses on structural similarity rather than strict temporal alignment.}\label{fig:DTW}
\end{figure}

\clearpage

\section{Comparing spread and error in HENS TC tracks}

\begin{figure}[H]
    \centering
    \includegraphics[width=0.6\linewidth]{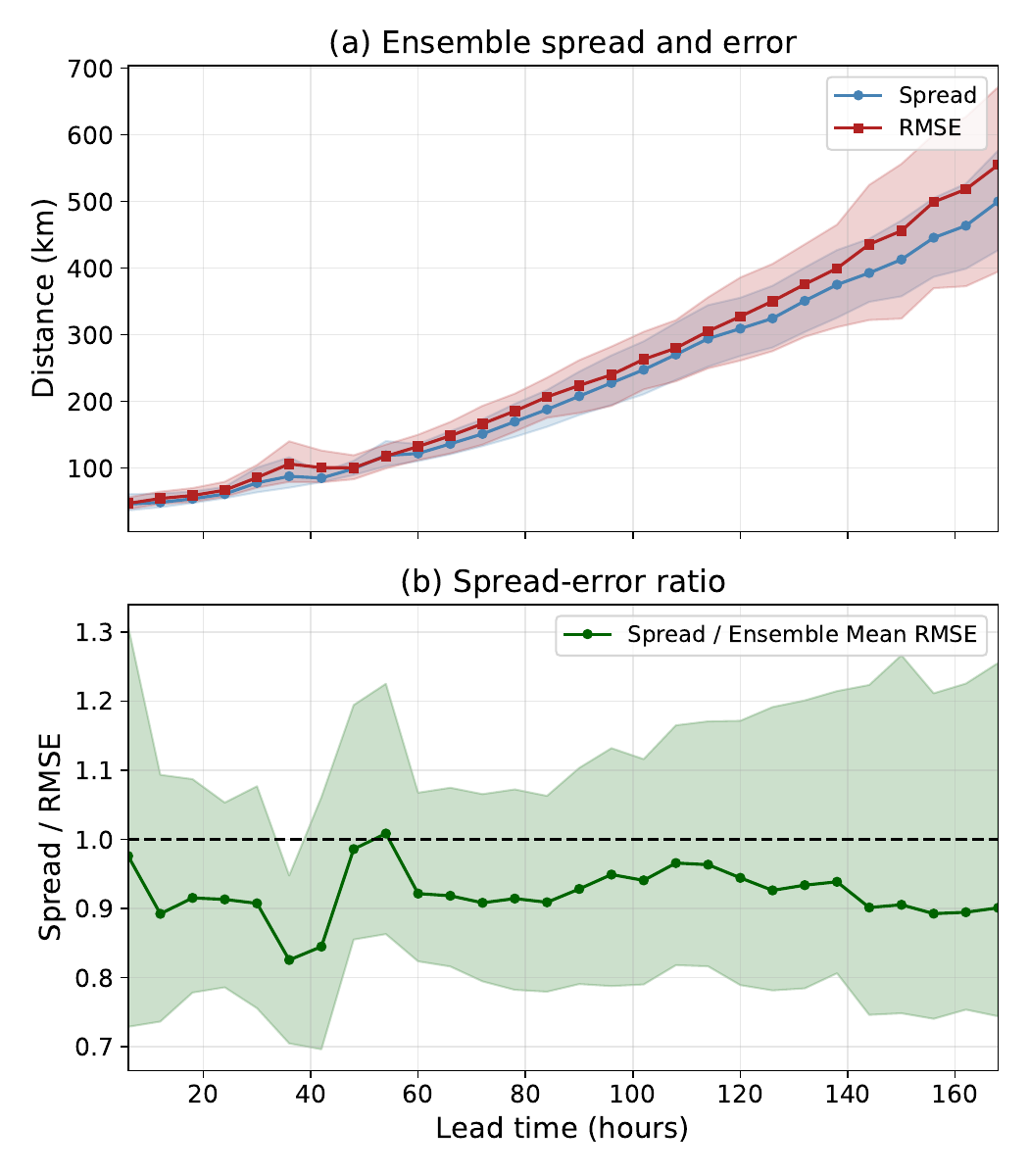}
    \caption{\textbf{Spread–error relationship of HENS tropical cyclone track forecasts.}
    (a)~Ensemble spread (blue; root-mean ensemble variance of member positions about the
    ensemble-mean track) and ensemble-mean position error (red; RMSE relative to the IBTrACS
    observed track), as a function of lead time, following \protect\cite{Fortin2014} (the square root is taken after pooling). (b)~The spread–error ratio; a value of 1 (dashed line) indicates a well-calibrated ensemble, below 1 underdispersion, and above 1 overdispersion. Statistics are pooled over all initialization dates and the named TCs of Table~A1, conditional on ensemble members that retain a tracked, matched cyclone at each lead time, and restricted to forecasts sampled by at least 100 members. Shaded bands are 95\% confidence intervals obtained by bootstrapping over storms.}
    \label{fig:spread_error}
\end{figure}

\clearpage 
\section{Calculating the diffusion exponents in IBTrACS}

\begin{figure}[H]
    \centering
    \includegraphics[width=0.5\linewidth]{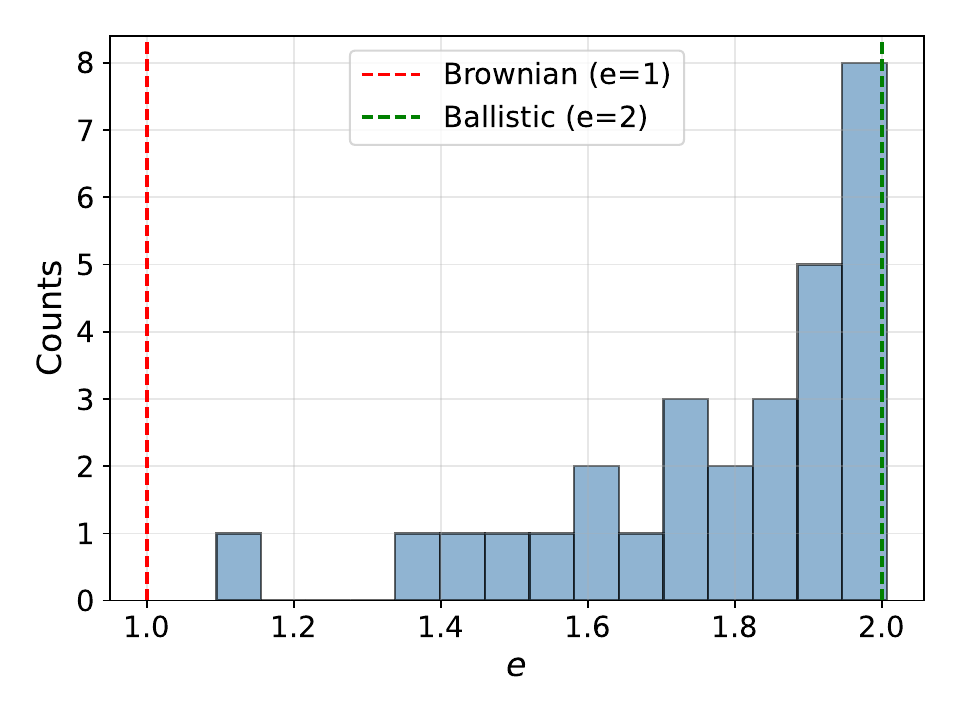}
\caption{\textbf{Distribution of anomalous diffusion exponents from IBTrACS observations.}  Based on the observed track of each TC in IBTrACS, $e$ is calculated for each TC, using Equations~3 and~4.  Due to the use of IBTrACS observations, instead of HENS, only one value of $e$ can be calculated for each storm, and the distribution across all TCs in summer 2023 is shown. Results are shown for the 33 TCs listed in Table~A1, with the exception of four TCs (Bret, Franklin, Hilary, and Idalia) that were excluded due to irregular time intervals in IBTrACS. Dashed lines indicate Brownian ($e=1$) and ballistic ($e=2$) diffusion limits.}
    \label{fig:ibtracs_msd_histogram}
\end{figure}

%



%
%

\newpage 
\section*{Open Research Section}
\label{sec:OpenResearch}

The associated data files generated used to calculate the TC tracks in HENS is available at \url{https://datadryad.org/dataset/doi:10.5061/dryad.ttdz08mbt} \cite{TCDryad}.  The code to conduct the analysis in this manuscript and generate the figures is available at \url{https://zenodo.org/records/21159051} \cite{TCCodeDryad}.

In order to generate the HENS dataset itself introduced in \cite{Mahesh2024hugeensemblesidesign, Mahesh2024hugeensemblespartii}, the code, datasets, and models are all stored at \url{https://doi.org/10.5061/dryad.2rbnzs80n}, via DataDryad \cite{HENSDryadCode}.  The code for HENS is integrated with Zenodo at the prior DOI. We include the code to train SFNO and conduct ensemble inference with bred vectors and multiple checkpoints.  We open-source the model weights of the trained SFNO checkpoints.  See the README of the DOI for information on how to use the codebase and for the permissive license associated with the code and data.  The code is available via the Lawrence Berkeley Lab BSD variant license, and the data is available with a CC0 license.

\begin{itemize}
    \item TempestExtremes for Emulated TCs: We use TempestExtremes at the GitHub repository \url{https://github.com/ClimateGlobalChange/tempestextremes} to identify and track emulated tropical cyclones in HENS ensemble data. Configuration files for running TempestExtremes on HENS are available at \url{https://zenodo.org/records/14939489}. 
\end{itemize}

\acknowledgments

This research was supported by the Director, Office of Science, Office of Biological and Environmental Research of the U.S. Department of Energy under Contract No. DE-AC02-05CH11231 and by the Regional and Global Model Analysis Program area within the Earth and Environmental Systems Modeling Program (WDC, AM). The research used resources of the National Energy Research Scientific Computing Center (NERSC), which is also supported by the Office of Science of the U.S. Department of Energy, under Contract No. DE-AC02-05CH11231.  The computation for this paper was supported in part by the DOE Advanced Scientific Computing Research (ASCR) Leadership Computing Challenge (ALCC) 2023-2024 award ‘Huge Ensembles of Weather Extremes using the Fourier Forecasting Neural Network’ to William Collins (LBNL) and the 2024-2025 award ‘Huge Ensembles of Weather Extremes using the Fourier Forecasting Neural Network’ to William Collins (LBNL).  The NERSC data lake is maintained by the PCMDI project and funded by the Earth and Environmental Sciences Division of the U.S. Department of Energy.

%
%

\raggedright
\bibliography{agusample}

%
%
%
%
%

\end{document}